\documentclass[12pt,draftcls]{IEEEtran}

\usepackage[mathcal]{euscript}
\usepackage{epsfig}
\usepackage{mathrsfs}
\usepackage{stmaryrd}
\usepackage{amsmath,amstext,amsfonts,amssymb}
\usepackage{epsfig,float}
\usepackage{graphicx}
\usepackage{cite}
\usepackage{psfrag}
\usepackage{subfigure}
\usepackage{url}
\usepackage{shadow,color,pifont,times,rotate}
\onecolumn
\DeclareMathAlphabet{\mathpzc}{OT1}{pzc}{m}{it}

\newcommand{\mb}{\mathbf}

\newcommand{\mc}{\mathcal}

\newtheorem{theorem}{Theorem}[section]
\newtheorem{lemma}[theorem]{Lemma}

\newtheorem{corollary}[theorem]{Corollary}

\renewcommand{\baselinestretch}{1.8}
\def\Mcb{M}

\def\H{{\mathbf H}}
\def\HH{\widehat{{\mathbf H}}}

\def\y{{\mathbf y}}
\def\s{{\mathbf x}}     %a cause de notations de Pablo

\def\W{{ W}}

\def\Esp{\mathbb{E}}

\title{On the Outage Capacity of a Practical Decoder\\ [-6mm] Accounting for Channel Estimation\\ [-6mm] Inaccuracies${\scriptsize ^{1}}$\footnote{$^1${\footnotesize The material in this paper was published in part at the International Symposium on Information Theory (ISIT07).}}}
\author{Pablo Piantanida$^*$, Sajad Sadough$^{*\dagger}$ and Pierre Duhamel$^*$\\[1mm]
{\normalsize $^*$Laboratoire des Signaux et Syst\`emes, CNRS/Sup\'{e}lec,\\[-3mm]
 F-91192 Gif-sur-Yvette, France \\[-3mm] 
Email:\{piantanida,pierre.duhamel\}@lss.supelec.fr \\[-3mm]
$^\dagger$Ecole Nationale Sup\'erieure de Techniques Avanc\'ees, 75015 Paris, France\\[-5mm]
Email: sajad.sadough@ensta.fr} \vspace{-2cm}}
\date{July, 2007}

\begin{document}
\maketitle
\begin{abstract} 
The optimal decoder achieving the outage capacity under imperfect channel estimation is investigated. First, by searching into the family of nearest neighbor decoders, which can be easily implemented on most practical coded modulation systems, we derive a decoding metric that minimizes the average of the transmission error probability over all channel estimation errors. Next, we specialize our general expression to obtain the corresponding decoding metric for fading MIMO channels. According to the notion of estimation-induced outage (EIO) capacity introduced in our previous work and assuming no channel state information (CSI) at the transmitter, we characterize maximal achievable information rates, using Gaussian codebooks, associated to the proposed decoder. In the case of uncorrelated Rayleigh fading, these achievable rates are compared to the rates achieved by the classical mismatched maximum-likelihood (ML) decoder and the ultimate limits given by the EIO capacity. Numerical results show that the derived metric provides significant gains for the considered scenario, in terms of achievable information rates and bit error rate (BER), in a bit interleaved coded modulation (BICM) framework, without introducing any additional decoding complexity. \vspace{-6mm}
\end{abstract}

\begin{keywords}\vspace{-.5cm}
Fading channels, Maximum likelihood estimation, Information rates, Decoding, MIMO systems.\vspace{-.6cm}
\end{keywords}

%###################################################################################'
\section{Introduction}\label{chII-sectionI}
%###################################################################################'

Consider a practical wireless communication system, where the receiver disposes only of noisy channel estimates that may in some circumstances be poor estimates, and these estimates are not available at the transmitter. This constraint constitutes a practical concern for the design of such communication systems that, in spite of their knowledge limitations, have to ensure communications with a prescribed quality of service (QoS). This QoS requires to guarantee transmissions with a given target information rate and small error probability, no matter which degree of accuracy estimation arises during the transmission. The described scenario addresses two important questions: (i) What are the theoretical limits of reliable transmission rates, using the best possible decoder in presence of imperfect channel state information at the receiver (CSIR) and (ii) how those limits can be achieved by using practical decoders in coded modulation systems ? Of course, these questions are strongly related to the notion of capacity that must take into account the above mentioned constraints. 

We have addressed in \cite{piantanida_it2007} the first question (i), for arbitrary memoryless channels, by introducing the notion of \emph{Estimation-Induced Outage Capacity} (EIO capacity). This novel notion characterizes the information-theoretic limits of such scenarios, where the transmitter and receiver strive to construct codes for ensuring the desired communication service, no matter which degree of accuracy estimation arises during the transmission. The explicit expression of this capacity allows one to evaluate the optimal trade-off between the maximal achievable outage rate (i.e. maximizing over all possible transmitter-receiver pairs) versus the outage probability $\gamma_{_{QoS}}$ (the QoS constraint). This can be used by a system designer to optimally share the available resources (e.g. power for transmission and training, the amount of training used, etc.), so that the communication requirements be satisfied. Nevertheless, the theoretical decoder used to achieve the latter capacity cannot be implemented on practical communication systems. 

The second question (ii) concerning the derivation of a practical decoder, which can achieve information rates close to the EIO capacity, is addressed in this paper. Classically, one replaces the exact channel by its estimate in the decoding metric. This is known as mismatched maximum-likehood (ML) decoding. However, this scheme is not appropriate in presence of channel estimation errors (CEE), at least if the estimation errors are large, i.e. for small number of training symbols \cite{taricco-biglieri-2005}. This problem has recently motivated a lot of work. In \cite{Ahmed-2004} and \cite{leke-1998} the authors analyze bit error rate (BER) performances of this mismatched decoder in the case of an orthogonal frequency division multiplexing (OFDM) system. References \cite{Garg-2005} considered a training-based MIMO system and showed that for compensating the performance degradation due to CEE, the number of receive antennas should be increased, which may become a limiting item for mobile applications. On the other hand, the performance of Bit Interleaved Coded Modulation (BICM) over fading MIMO channels with perfect CSI was studied for instance, in \cite{caire-tarico-1998}, \cite{zehavi} and \cite{Li-2002}. Cavers in \cite{Cavers-1991}, derived a tight upper bound on the symbol error rate of pilot symbol assisted modulation (PSAM) for a $16$-QAM constellation. A similar investigation was carried out in \cite{Huang-2003} showing that for iterative decoding of BICM at low SNR, the quality of channel estimates is too poor for being used in the mismatched ML decoder. 

As an alternative to the aforementioned decoder, Tarokh {\it et al.} in \cite{Tarokh-1999} and Taricco and Biglieri in \cite{taricco-biglieri-2005}, proposed an improved ML detection metric and applied it to a space-time coded MIMO system, where they showed the superiority of this metric in terms of BER. Interestly enough, this decoding metric can be formally derived as a special case of the general framework presented in this paper. So far, most of the research in the field were focused on evaluating the performances of mismatched decoders in terms of BER (cf. \cite{Divsalar-1979}), but still not providing an answer to the question (ii). In \cite{Weingarten-2004}, the authors investigate achievable rates of a weighting nearest-neighbor decoder for multiple-antenna channel. Moreover, in \cite{LS02} and \cite{piantanida_it2007}, authors show that the achievable rates using the mismatched ML decoding are largely sub-optimal (at least for a limited number of training symbols) compared to the ultimate limits given by the EIO capacity. In this paper, according to the notion of EIO capacity, we investigate the maximal achievable information rate with Gaussian codebooks of the improved decoder in \cite{Tarokh-1999,taricco-biglieri-2005}. Furthermore, it can be shown that this decoder achieves the capacity of a composite (more noisy) channel.

This paper is organized as follows. Section \ref{chII-sectionII}, briefly reviews our notion of capacity. Then, we search into the family of decoders that can be easily implemented on most practical coded modulation systems to derive the general expression of the decoder. This decoder minimizes the average of the transmission error probability over all CEE. We accomplish this by exploiting the availability of the statistic characterizing the quality of channel estimates, i.e., the {\it a posteriori} probability density function (pdf) of the unknown (true) channel conditioned on its estimate. Section \ref{chII-sectionIII} describes the fading MIMO model. In section \ref{chII-sectionIV}, we specialize our expression of the decoding metric for the case of MIMO channels and use this for iterative decoding of MIMO-BICM. In section \ref{chII-sectionV}, we compute achievable information rates of a receiver using the proposed decoder and compare these to the EIO capacity and the achievable rates of the classical mismatched approach. Section \ref{chII-sectionVI} illustrates via simulations, conducted over uncorrelated Rayleigh fading, the performance of the improved decoder in terms of achievable outage rates and BER, compared to those provided by the mismatched ML decoding. %Finally, section VII concludes the paper. 

Notational conventions are as follows. Upper and lower case bold symbols are used to denote matrices and vectors; $\mathbb{I}_{M}$ represents an $(M \times M)$ identity matrix; $\mathbb{E}_\mb{X}\{\cdot\}$ refers to expectation with respect to the random vector $\mb{X}$; $|\cdot|$ and $\|\cdot\|_F$ denote matrix determinant and Frobenius norm, respectively; $(\cdot)^T$ and $(\cdot)^\dag$ denote vector transpose and Hermitian transpose, respectively. \vspace{-2mm}

%###################################################################################'
\section{Decoding under Imperfect Channel Estimation}\label{chII-sectionII}
%###################################################################################'

Throughout this section we focus on deriving a practical decoder for general memoryless channels that achieves information rates close to the EIO capacity (the ultimate bound). %Before this, let us first review the considered DMC model with input $x \in\mathscr{X}$ and output $y\in \mathscr{Y}$, and the above mentioned notion of capacity.  \vspace{-3mm}

\subsection{Communication Model Under Channel Uncertainty} 

A specific instance of the memoryless channel is characterized by a transition probability $W(y|x,\theta)\in\mc{W}_{\Theta}$ with an unknown channel state $\theta$, over input and output alphabets $ \mathscr{X}, \mathscr{Y}$. Here, $\mc{W}_{\Theta}=\big\{W(\cdot|x,\theta)\!: \, x\in \mathscr{X},\,\theta\in \Theta\big\}$ is a family of conditional pdf parameterized by the vector of parameters $\theta\in \Theta\subseteq \mathbb{C}^d$, where $d$ denotes the number of parameters. Throughout the paper we assume that the channel state, which neither the transmitter nor the receiver know exactly, remains constant within blocks of symbols, related to the product of the coherence time and the coherence bandwidth of a wireless channel, and these states for different blocks are i.i.d.~$\theta\sim\psi(\theta)$ (e.g. block Rayleigh fading). The transmitter does not know 
$\theta$ and the receiver only knows an estimate $\hat{\theta}$ and a \emph{characterization of the estimator performance} in terms of the conditional pdf $\psi(\theta|\hat{\theta})$ (obtained by using $\mc{W}_{\Theta}$, the estimation function and $\psi(\theta)$). A decoder using $\hat{\theta}$, instead of $\theta$, obviously might not support an information rate $R$ (even small rates might not be supported if $\hat{\theta}$ and $\theta$ are strongly different). Consequently, outage events induced by CEE will occur with a certain probability $\gamma_{_{QoS}}$. The scenario underlying these assumptions is motivated by current wireless systems, where the coherence time for mobile receivers may be too short to permit reliable estimation of the fading coefficients and in spite of this fact,  the desired communication service must be guaranteed. This leads to the following notion of capacity. \vspace{-4mm}

%-------------------------------------------------------------------------------------------------------------------
\subsection{A Brief Review of EIO Capacity} \label{chII-brief_review_capa}
%-------------------------------------------------------------------------------------------------------------------

A message $m\in\mc{M}=\{1,\dots,\lfloor\exp(nR)\rfloor \}$ is transmitted using a pair $(\varphi,\phi)$ of mappings, where $\varphi: \mc{M}     \mapsto   \mathscr{X}^n$ is the encoder, and $\phi: \mathscr{Y}^n\times  \Theta  \mapsto   \mc{M}$ is the decoder (that utilizes $\hat{\theta}$). The random rate, which depends on the unknown channel realization $\theta$ through its probability of error, is given by $n^{-1}\log \Mcb_{\theta,\hat{\theta}}$. The maximum error probability (over all messages) 
\begin{equation}
e_{\max}^{(n)}(\varphi,\phi,\hat{\theta};\theta)=\max_{m\in\mc{M}} \int_{\{\mathbf{y}\in\mathscr{Y}^n:\phi(\mathbf{y}, \hat{\theta})\neq m\}}\!\!\!\!\!\!\!\!\!\!\! dW^n\big(\mathbf{y}|\varphi(m),\theta\big),\label{chII-eq-error}
%\max_{m\in\mc{M}}W^n\big(\{\phi(\mathbf{y}, \hat{\theta})\neq m\}\big|\varphi(m),\theta\big),\label{chII-eq-error}
\end{equation}
where $\mb{y}=(y_1,\dots,y_n)$. For a given channel estimate $\hat{\theta}$, and $0<\epsilon,\gamma_{_{QoS}}< 1$, an outage rate $R\geq 0$ is $(\epsilon,\gamma_{_{QoS}})$-achievable if for every $\delta>0$ and every sufficiently large $n$ there exists a sequence of length-$n$ block codes such that the rate satisfies the quality of service
\begin{equation}
\Pr\left(\Lambda_\epsilon(R,\hat{\theta}) \big | \hat{\theta} \right)=\int_{\Lambda_\epsilon(R,\hat{\theta})}\!\!\!  d\psi(\theta|\hat{\theta}) \geq 1-\gamma_{_{QoS}},\label{chII-eq-proba}
\end{equation}
where $\Lambda_\epsilon(R,\hat{\theta})=\big\{\theta\in \Delta_{\epsilon}^{(n)}: n^{-1}\log \Mcb_{\theta,\hat{\theta}}\,\geq\, R-\delta \big\}$ stands for the set of all channel states allowing for the desired transmission rate $R$, and $\Delta_{\epsilon}^{(n)} = \big\{\theta\in\Theta\!: \,e_{\max}^{(n)}(\varphi,\phi,\hat{\theta};\theta)\leq\epsilon\big\}$ is the set of all channel states allowing for reliable decoding (arbitrary small error probability). This definition requires that maximum error probabilities larger than $\epsilon$ occur with probability less than $\gamma_{_{QoS}}$. The practical advantage of such definition is that for $(1-\gamma_{_{QoS}})$\% of channel estimates, the transmitter and receiver strive to construct codes for ensuring the desired communication service. The EIO capacity is then defined as the largest $(\epsilon,\gamma_{_{QoS}})$-achievable rate, for an outage probability $\gamma_{_{QoS}}$ and a given channel estimate $\hat{\theta}$, as
\begin{equation}
C(\gamma_{_{QoS}},\hat{\theta})=\lim\limits_{\epsilon\downarrow  0}\sup\limits_{\varphi,\phi}\Big\{R\geq 0:\,  \Pr\big(\Lambda_\epsilon(R,\hat{\theta})| \hat{\theta}\big)\geq1-\gamma_{_{QoS}}  \Big\},\label{chII-eq-max}
\end{equation}
where the maximization is taken over all encoder and decoder pairs. In \cite{piantanida_it2007}, we proved the following coding Theorem that provides an explicit way to evaluate the maximal outage rate \eqref{chII-eq-max} versus outage probability $\gamma_{_{QoS}}$ for an estimate $\hat{\theta}$, characterized by $\psi(\theta|\hat{\theta})$. %This can be written in a compact form as follows.
\begin{theorem}\label{chII-theo-capacity}
Given an outage probability $0\leq\gamma_{_{QoS}}<1$, the EIO capacity is given by
\begin{equation}
\label{chII-outage-cap}
C(\gamma_{_{QoS}},\hat{\theta})=\max\limits_{P\in\mathscr{P}_\Gamma(\mathscr{X})}\sup\limits_{\Lambda\subset \Theta:\,\,\Pr(\Lambda|\hat{\theta})\geq 1-\gamma_{_{QoS}}}\!\!\!\!\!\!\inf\limits_{\theta\in\Lambda}I\big(P,W(\cdot|\cdot,\theta)\big),
\end{equation}
where $I(\cdot)$ denotes the mutual information of the channel $W(y|x,\theta)$ and $\mathscr{P}_\Gamma(\mathscr{X})$ is the set of input distributions that does not depend on $\hat{\theta}$, satisfying the input constraint $\int g(x) dP(x) \leq\Gamma$ for a nonnegative cost function $g:\mathscr{X}\rightarrow [0,\infty)$. \vspace{.1mm}
\end{theorem}

%$$
%I\big(P,W(\cdot|\cdot,\theta)\big) = \sum_{x\in\mc{X}}\sum_{y\in\mc{Y}}
%P(x)W(y|x,\theta)\log \frac{W(y|x,\theta)}{ Q(y|\theta) },
%$$
%with $ Q(y|\theta) = \sum_{x\in\mc{X}} P(x)W(y|x,\theta)$. We emphasize that the supremum in \eqref{chII-qe-capacity} is taken over all subsets $\Lambda$ of $\Theta$ that have (conditional) probability at least $1-\gamma_{_{QoS}}$. 

The existence of a decoder $\phi$ in \eqref{chII-eq-max} achieving the capacity \eqref{chII-outage-cap} is proved using a random-coding argument, based on the well-known method of typical sequences \cite{csiszar-book}. Nevertheless, this decoder cannot be implemented on practical communication systems. 

%-------------------------------------------------------------------------------------------------------------------------------
\subsection{Derivation of a Practical Decoder Using Channel Estimation Accuracy}
%-------------------------------------------------------------------------------------------------------------------------------

We now consider the problem of deriving a practical decoder that achieves the capacity \eqref{chII-outage-cap}. Assume that we restrict the searching of decoding functions $\phi$, maximizing \eqref{chII-eq-max}, to the class of additive decoding metrics, which can be implemented on realistic systems. This means that for a given channel output $\mb{y}=(y_1,\dots,y_n)$, we set the decoding function
\begin{equation}
\phi_{\mc{D}}(\mb{y},\hat{\theta})=\arg\min\limits_{m\in \mc{M}}\mc{D}^n\big(\varphi(m),\mb{y}|\hat{\theta}\big), \label{chII-decoder-metric}
\end{equation}
where $\mc{D}^n\big(\mb{x},\mb{y}|\hat{\theta}\big)=n^{-1}\sum_{i=1}^n \mc{D}\big(x_i,y_i|\hat{\theta}\big)$ and $\mc{D}: \mathscr{X}\times \mathscr{Y}\times \Theta  \mapsto \mathbb{R}_{\geq 0}$ is an arbitrary per-letter additive metric.  Consequently, the maximization in \eqref{chII-eq-max} is actually equivalent to maximizing over all decoding  metrics $\mc{D}$. Note, however, that this restriction does not necessarily lead to an optimal decoder achieving the capacity. %However, these additive metrics can be easily implemented on most practical coded modulation systems, e.g. BICM.

\emph{Problem statement:} In order to find the optimal decoding metric $\mc{D}$ maximizing the outage rates in \eqref{chII-eq-max}, for a given outage probability $\gamma_{_{QoS}}$ and channel estimate $\hat{\theta}$, it is necessary to look at the intrinsic properties of the capacity definition. Observe that the size of the set of all channel states allowing for reliable decoding $\Delta_{\epsilon}^{(n)}$ is determined by the decoding function $\phi$. The maximal achievable rate $R$, constrained to the outage probability \eqref{chII-eq-proba}, is thus limited by this size. Hence, for a given decoder $\phi$, there exists an optimal set $\Lambda^*_\epsilon \subseteq \Delta_{\epsilon}^{(n)}$ of channel states with conditional probability larger than $1-\gamma_{_{QoS}}$, providing the largest achievable rate, which follows as the minimal instantaneous rate for the worst $\theta \in\Lambda^*_\epsilon$. The optimal set $\Lambda^*_\epsilon$ is equal to the set $\Lambda^*$ maximizing the expression \eqref{chII-outage-cap}. Hence, an optimal decoding metric must guarantee minimum error probability \eqref{chII-eq-error} for every $\theta \in\Lambda^*$. 

The computation of such a metric becomes very difficult (not necessary feasible by using the class of decoders in \eqref{chII-decoder-metric}), since the maximization in \eqref{chII-eq-max} by using $\phi_\mc{D}$ is not an explicit function of $\mc{D}$. However, it is interesting to note, that if the set $\Lambda^*$ defines a compact and convex set of channels $\mc{W}_{\Lambda^*}$, then the optimal decoding metric can be chosen as the ML decoder $\mc{D}^*(x,y|\hat{\theta})=-\log W(y|x,\theta^*)$, where $\theta^*$ is the channel state minimizing the mutual information in \eqref{chII-outage-cap}. The receiver can thus be a ML receiver with respect to the worst channel in the family \cite{csiszar-1995}. However, in most practical cases, the channel states are represented by vectors of complex coefficients that do not lead to convex sets of channels. 

\emph{Optimal decoder for composite channels:} Instead of trying to find an optimal decoding metric minimizing the error probability \eqref{chII-eq-error} for every $\theta \in\Lambda^*$, we propose to look at the decoding metric minimizing the average of the transmission error probability over all CEE. This means,\vspace{-1mm}
\begin{equation}
\mc{D}_\mc{M}=\arg\min\limits_{\mc{D}} \int_\Theta e_{\max}^{(n)}(\varphi,\phi_{\mc{D}},\hat{\theta};\theta) d\psi(\theta|\hat{\theta}),\label{chII-eq_average_error}\vspace{-1mm}
\end{equation}
where $e_{\max}^{(n)}$ is obtained by replacing \eqref{chII-decoder-metric} in \eqref{chII-eq-error}. Since the channel $W$ is memoryless, the average of error probability in \eqref{chII-eq_average_error} can be written as the error probability of a composite (more noisy) channel $\widetilde{W}(y|x,\hat{\theta})$. This channel follows as the average of the unknown channel $W$ over all CEE given the estimate $\hat{\theta}$. Then, by taking the logarithm of this channel we obtain its ML decoder, which minimizes (for $n$ sufficiently large) the error probability in \eqref{chII-eq_average_error}. Actually, by following an analogy with the proof in \cite{csiszar-1995}, it can be shown that \vspace{-1mm}
\begin{eqnarray}
\mc{D}_\mc{M}(x,y|\hat{\theta})=-\log \widetilde{W}(y|x,\hat{\theta})\,\,\,\, \textrm{with}\,\, \,\, \widetilde{W}(y|x,\hat{\theta})=\int_\Theta W(y|x,\theta) d\psi(\theta|\hat{\theta}).\label{chII-metric-definition}
\end{eqnarray}

\emph{Remark:}  We emphasize that this decoder cannot guarantee small error probabilities for every channel state $\theta\in\Lambda^*$, and consequently it only achieves a lower bound of the EIO capacity \eqref{chII-outage-cap}. Nevertheless, this archives the capacity of the composite channel. The remaining question to answer is how much lower are the achievable outage rates using the metric \eqref{chII-metric-definition}, comparing to the theoretical decoder achieving the EIO capacity. In section \ref{chII-sectionV}, we evaluate \eqref{chII-metric-definition} and its achievable information rates for the fading MIMO channel with no CSI at the transmitter. \vspace{-2mm}

%########################################################################################
\section{System Model}\label{chII-sectionIII}\vspace{-2mm}
%########################################################################################
%-----------------------------------------------------------------------------------------------
\subsection{Fading MIMO Channel}
\begin{figure}[!htb] 
\centering
\includegraphics[width=0.8\textwidth,height=0.18\textheight]{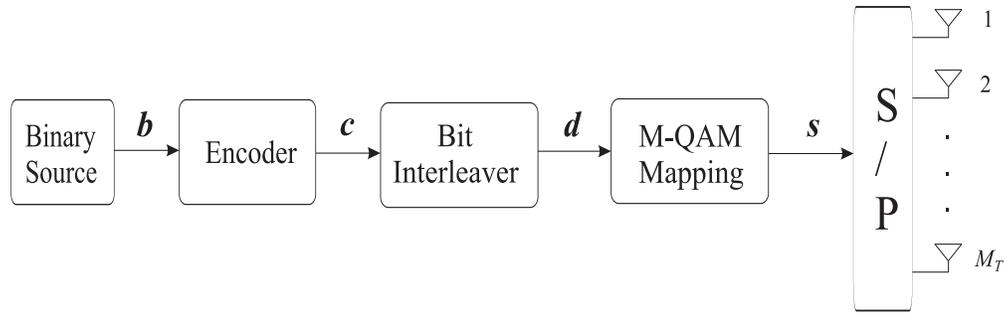} 
\caption{Block diagram of MIMO-BICM transmission scheme.} \label{chII-fig_1}
\end{figure}

\vspace{-1mm}
%-----------------------------------------------------------------------------------------------
We consider a single-user MIMO system with $M_T$ transmit and $M_R$ receiver antennas transmitting over a frequency non-selective channel and refer to it as a MIMO channel. Fig. \ref{chII-fig_1} depicts the BICM coding scheme used at the transmitter. The binary data sequence $\mb{b}$ is encoded by a non-recursive and non-systematic convolutional (NRNSC) code, before being interleaved by a quasi-random interleaver. The output bits $\mb{d}$ are gathered in subsequences of $B$ bits and mapped to complex M-QAM $(M = 2^B)$ vector symbols $\mb{x}$ with average power $\displaystyle{\frac{tr(\mb{x}\mb{x}^\dag )}{M_T}}= \bar{P}$. We also send some pilot symbols at the beginning of each data frame for channel estimation. The symbols of a frame are then multiplexed for being transmitted through $M_T$ antennas. Assuming a frame of  $L$ transmitted symbols associated to each channel matrix $\H_k$, the received signal vector $\mb{y}_k$ of dimension $(M_R \times 1)$ is given by 
\begin{equation}
\mb{y}_k = \H_k \mb{x}_k + \mb{z}_k,\:\:\: k = 1,\dots,L, \label{chII-MIMO_model} 
\end{equation}
where $\mb{x}_k $ is the $(M_T \times 1)$ vector of transmitted symbols, referred to as a compound symbol. Here, the entries of the random matrix $\H_k$ are independent identically distributed (i.i.d.) Zero-Mean Circularly Symmetric Complex Gaussian (ZMCSCG) random variables. Thus, the channel state $\theta=\H_k$ is distributed as $\H_k \sim \psi_H(\mb{H})=C\mc{N}\big(\mb{0}, \mathbb{I}_{M_T}\otimes \mb{\Sigma_{H}}\big)$
\begin{equation}
C\mc{N}\big(\mb{0}, \mathbb{I}_{M_T}\otimes \mb{\Sigma_{H}}\big)=\frac{1}{\pi^{M_R M_T}|\Sigma_{H} |^{M_T}}\exp\Big[-tr \big( \mb{H}\mb{\Sigma_{H}}^{-1} \mb{H}^\dag  \big)\Big],\label{chII-pdf_MIMO}
\end{equation}
where $\mb{\Sigma_{H}}$ is the Hermitian covariance matrix of the columns of $\mb{H}$ (assumed to be the same for all columns), i.e., $\mb{\Sigma_{H}}=\sigma_{H}^2\mathbb{I}_{M_R}$. The noise vector $\mb{z}_{k}\in \mathbb{C}^{M_R\times 1}$ consists of ZMCSCG random vector with covariance matrix $\mb{\Sigma_{0}}=\sigma_{Z}^2\mathbb{I}_{M_R}$. Both $\mb{H}_k$ and $\mb{z}_k$ are assumed ergodic and stationary random processes, and the channel matrix $\mb{H}_k$ is independent of $\mb{x}_k$ and $\mb{z}_k$. \vspace{-4mm}

%-----------------------------------------------------------------------------------------------
\subsection{Pilot Based Channel Estimation}
%-----------------------------------------------------------------------------------------------

Assuming that the channel matrix is time-invariant over an entire frame, channel estimation is usually performed on the basis of known training (pilot) symbols transmitted at the beginning of each frame. The transmitter, before sending the data $\mb{x}_k$, sends a training sequence of $N$ vectors $\mb{X}_T=(\mb{x}_{T,1},\dots,\mb{x}_{T,N})$. According to the observation of the channel model \eqref{chII-MIMO_model}, this sequence is affected by the channel matrix $\mb{H}_k$, allowing the receiver to observe separately $\mb{Y}_{T,k}=\mb{H}_k\,\mb{X}_{T,k}+\mb{Z}_{T,k}$, where $\mb{Z}_{T,k}$ is the noise matrix affecting the transmission of training symbols. We assume that the coherence time is much longer than the training time and the average energy of the training symbols is $\bar{P}_T=\frac{1}{N M_T}tr\big(\mb{X}_T \mb{X}_T^\dag\big)$.

We focus on the estimation of $\mb{H}_k$, from the observed signals $\mb{Y}_{T,k}$ and $\mb{X}_{T,k}$. In the ML sense this estimate is obtained by minimizing $\|\mb{Y}_{T,k}- \mb{H}_k\mb{X}_T \|^2$ with respect to $\mb{H}_k$. This yields $\widehat{\mb{H}}_{\textrm{ML},k}= \mb{Y}_{T,k}\mb{X}_T^\dag \big(\mb{X}_T \mb{X}_T^\dag\big)^{-1}=\mb{H}_k+\mc{E}_k$, where $\mc{E}_k = \mb{Z}_{T,k} \mb{X}_T^\dag \big(\mb{X}_T \mb{X}_T^\dag\big)^{-1}$ denotes the estimation error matrix. For simplicity, we assume orthogonal training sequences, for which we must have $N\geq M_T$, and consequently the matrix error becomes decorrelated.  
%Since to estimate the $M_R\times M_T$ channel matrix, we need at least $M_R M_T$ independent measurements, and each symbol time yields $M_R$ samples at the receiver, we must have $N\geq M_T$. 
Thus, matrix $\mb{X}_T$ must be full rank $M_T$ and thus $\mb{X}_T \mb{X}_T^\dag$ must be nonsingular with orthogonal rows and such that $\mb{X}_T\mb{X}_T^\dag=N P_T \mathbb{I}_{M_T}$. Next, denoting $\underline{\mc{E}}_j$ the $j$th column of the error matrix $\mc{E}$, we can write $\mb{\Sigma_{\mc{E}}}=\mathbb{E}_{\mc{E}}\big\{\underline{\mc{E}}_j \underline{\mc{E}}_j^\dag \big\}=\textrm{SNR}^{-1}_{T}\mathbb{I}_{M_R}$ with  $\textrm{SNR}_{T}=\displaystyle{\frac{N P_T}{\sigma^2_{Z}}}$, yielding a white error matrix, i.e. the entries of $\mc{E}$ are i.i.d. ZMCSCG random variables with variance $\sigma_{\mc{E}}^2=\textrm{SNR}^{-1}_{T}$. Thus, for each frame, the conditional pdf of $\hat{\theta}=\widehat{\mb{H}}_{\textrm{ML}}$ given $\theta=\mb{H}$ is the complex normal matrix pdf
\vspace{-1mm}
\begin{equation}
\psi_{\widehat{H}_{\textrm{ML}}|H}(\widehat{\mb{H}}_{\textrm{ML}}|\mb{H})=C\mc{N}\big(\mb{H}, \mathbb{I}_{M_T}\otimes \mb{\Sigma_{\mc{E}}}\big).
\label{chII-apriori-ML-pdf}
\end{equation}

%########################################################################################
\section{Metric Computation and Iterative Decoding of BICM}\label{chII-sectionIV}
%########################################################################################

In this section, we specialize the expression \eqref{chII-metric-definition} to derive the decoding metric for MIMO channels \eqref{chII-MIMO_model} and then we consider MIMO-BICM decoding with the derived metric. 

%---------------------------------------------------------------------------------------------------------------------------
\subsection{Mismatched ML Decoder}
%---------------------------------------------------------------------------------------------------------------------------
The classical mismatched ML decoder consists of the likelihood function of the channel pdf using the channel estimate $\HH_{\textrm{ML}}$. This leads to the following Euclidean distance
\begin{equation}
\mc{D}_{\textrm{ML}}\big( \mb{x}, \mb{y}|  \HH_{\textrm{ML}} \big)=-\log W(\mb{y}|\mb{x},\HH_{\textrm{ML}})=\|\mb{y}-\widehat{\mb{H}}_{\textrm{ML}} \mb{x}\|^2+\textrm{const}.\label{chII-mismached_ML} 
\end{equation}
%---------------------------------------------------------------------------------------------------------------------------
\subsection{Metric Computation}  
%---------------------------------------------------------------------------------------------------------------------------
We now specialize the expression \eqref{chII-metric-definition} in the case of a MIMO channel \eqref{chII-MIMO_model}. To this end, we need to derive the pdf $\psi_{H|\widehat{H}_\textrm{ML}}(\H |\HH_{\textrm{ML}})$, which can be obtained by using the pdf \eqref{chII-apriori-ML-pdf} and \eqref{chII-pdf_MIMO} (see Appendix \ref{chII-Appendix_I}). The corresponding pdf is: 
\begin{equation}
\psi_{H | \widehat{H}_{\textrm{ML}}} (\H | \HH_{\textrm{ML}})=C\mc{N}\big(\mb{\Sigma}_\Delta \HH_{\textrm{ML}}, \mathbb{I}_{M_T}\otimes \mb{\Sigma}_\Delta\mb{\Sigma_{\mc{E}}}\big),
\label{chII-aposteriori-ML-MIMO}
\end{equation}
where $\mb{\Sigma}_\Delta=\mb{\Sigma_H}(\mb{\Sigma_{\mc{E}}}+\mb{\Sigma_H})^{-1}=\mathbb{I}_{M_R} \delta$ and $\delta=\displaystyle{\frac{\textrm{SNR}_{T}\sigma^2_{H}}{\textrm{SNR}_{T}\sigma^2_{H}+1}}$. The availability of the distribution \eqref{chII-aposteriori-ML-MIMO} characterizing the CEE is the key feature of pilot assisted channel estimation. Then, by averaging the channel $W(\mb{y}|\mb{x},\H)$ over all CEE, using the pdf \eqref{chII-aposteriori-ML-MIMO}, and after some algebra we obtain  the composite channel (cf. Appendix \ref{chII-Appendix_I}) 
\begin{equation}\label{chII-composite-channel}
\widetilde{W}(\mb{y}|\mb{x},\HH_{\textrm{ML}})=C\mc{N}\big( \delta\HH_{\textrm{ML}}\mb{x}, \mb{\Sigma_{0}} + \delta \mb{\Sigma_{\mc{E}}}   \| \mb{x} \|^2     \big). 
\end{equation}
Finally, from \eqref{chII-composite-channel} the optimal decoding metric for the MIMO channel \eqref{chII-MIMO_model} reduces to:
\begin{equation}
\mc{D}_{\mc{M}}^{\textrm{MIMO}}\big( \mb{x}, \mb{y}|  \HH_{\textrm{ML}} \big)=M_R \log ( \sigma^2_Z+\delta \sigma_{\mc{E}}^2 \| \mb{x}\|^2)+\displaystyle{ \frac{\| \mb{y}- \delta \HH_{\textrm{ML}} \mb{x}  \|^2}{\sigma^2_Z+\delta \sigma_{\mc{E}}^2 \| \mb{x}\|^2}}. \label{chII-final_metric_MIMO}
\end{equation}
This metric coincides with that proposed for space-time decoding, from independent results in \cite{taricco-biglieri-2005}. We note that under near perfect CSI, obtained when $N\rightarrow \infty$,
\begin{equation}
\lim\limits_{N\rightarrow \infty}  \frac{\mc{D}_{\mc{M}}^{\textrm{MIMO}}\big( \mb{x}, \mb{y}|  \HH_{\textrm{ML}} \big)}{\mc{D}_{\textrm{ML}}\big( \mb{x}, \mb{y}|  \HH_{\textrm{ML}} \big)}=1, \,\,\,\,\,\textrm{\emph{almost surely.} }\label{chII-limit_metric}
\end{equation}
Consequently, we have the expected result that the metric \eqref{chII-final_metric_MIMO} tends to the classical mismatched ML decoding metric \eqref{chII-mismached_ML}, when the estimation error $\sigma_{\mc{E}}^2 \rightarrow 0$.  

%While we have focused on MIMO, the optimal decoding metric for each sub-currier of MB-OFDM channels is given by 
%\begin{equation}
%\mc{D}_{\mc{M},k}^{\textrm{OFDM}}\big(x_k,y_k| \widehat{H}_{\textrm{ML},k} \big)=\log ( \sigma^2_Z+\delta_k \sigma_{\mc{E}}^2 \| x_k \|^2)+\displaystyle{ \frac{\| y_k- \delta_k \widehat{H}_{\textrm{ML},k} x_k  \|^2}{\sigma^2_Z+\delta_k \sigma_{\mc{E}}^2 \| x_k\|^2}}, \label{chII-final_metric_OFDM}
%\end{equation}
%where $\delta_k=\frac{\textrm{SNR}_{T}\sigma^2_{h,k}}{\textrm{SNR}_{T}\sigma^2_{h,k}+1}$.
%---------------------------------------------------------------------------------------------------------------------------
\subsection{Receiver Structure}
\begin{figure}[!htb] 
%\vspace{-3mm}
\centering
\includegraphics[width=0.8\textwidth,height=0.25\textheight]{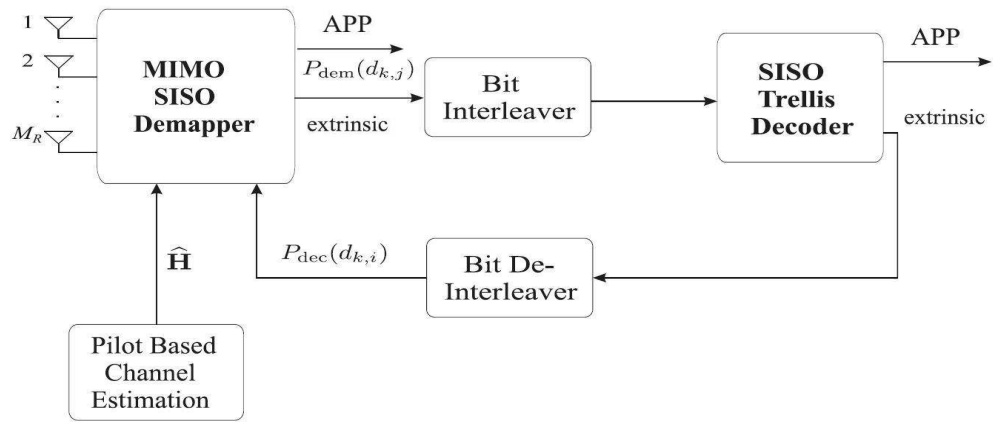} 
\caption{Block digram of MIMO-BICM receiver.} \label{chII-fig_3}
\end{figure}

%---------------------------------------------------------------------------------------------------------------------------
The problem of decoding MIMO-BICM has been addressed in \cite{boutros00} under the assumption of perfect CSIR. Here we consider the same problem with CEE, for which we use the metric \eqref{chII-final_metric_MIMO} in the iterative decoding process of BICM. Basically, the receiver consists of the combination of two sub-blocks operating successively. The block diagram of the transmitter and the receiver are shown in Fig. \ref{chII-fig_1} and Fig. \ref{chII-fig_3}, respectively. The first sub-block, referred to as soft symbol to bit MIMO demapper, produces bit metrics (probabilities) from the input symbols and the second one is a soft-input soft-output (SISO) trellis decoder. Each sub-block can take advantage of the {\it a posteriori} (APP) provided by the other sub-block as an a priori information. Here, SISO decoding is performed using the well known forward-backward algorithm \cite{bcjr}. We recall the formulation of the soft MIMO detector.

Suppose first the case where the channel matrix $\H$ is perfectly known at the receiver. The MIMO demapper provides at its output the extrinsic probabilities on coded and interleaved bits $\mb{d}$. Let $d_{k,j}$, $j = 1, ...,B M_T$, be the interleaved bits corresponding to the $k$-th compound symbol $\mb{x}_k \in Q$ where the cardinality of $Q$ is equal to $2^{BM_T}$. The extrinsic probability $P_{\rm dem}(d_{k,j})$ of the bit $d_{k,j}$ (bit metrics) at the MIMO demapper output is calculated as 
\begin{equation}\label{chII-eq:mimoRx1}
P_{\rm dem}(d_{k,j} = 1) = K \sum_{\stackrel{\mb{x}_k\in Q} {d_{k,j}=1}}  \prod_{\stackrel{i=1}{i\neq j}}^{B M_T} P_{\rm dec}(d_{k,i})\exp\big[-\mc{D}(\mb{x}_k,\mb{y}_k| \H_k) \big],
\end{equation}
where $\mc{D}(\mb{x}_k,\mb{y}_k| \H_k)=-\log \W(\y_k|\mb{x}_k,\H_k)$ and $K$ is the normalization factor satisfying $P_{\rm dem}(d_{k,j} = 1) + P_{\rm dem}(d_{k,j} = 0) = 1$ and $P_{\rm dec}(d_{k,i})$ is the {\it extrinsic} information coming from the SISO decoder. The summation in \eqref{chII-eq:mimoRx1} is taken over the product of the channel likelihood given a compound symbol $\mb{x}_k$, and the {\it a priori} probability on this symbol (the term $\prod P_{\rm dec}$) fed back from the SISO decoder at the previous iteration. Concerning this latter term, the {\it a priori} probability of the bit $d_{k,j}$ itself has been excluded, so as to let the exchange of extrinsic information between the channel decoder and the MIMO demapper. Also, note that this term assumes independent coded bits $d_{k,i}$, which is a valid approximation for random interleaving of large size. At the first iteration we set $P_{\rm dec}(d_{k,i})=1/2$ (there is no {\it a priori} information).

Note that by replacing the unknown channel in \eqref{chII-eq:mimoRx1} by its channel estimate $\HH_k$, we obtain the mismatched ML decoder \eqref{chII-mismached_ML}. The proposed decoder follows by introducing the metric given by $\mc{D}_{\mc{M}}^{\textrm{MIMO}}(\mb{x}_k,\mb{y}_k| \HH_k)$ in \eqref{chII-eq:mimoRx1}, yielding to the same equation with the appropriate constant $K$.  

%###################################################################################'
\section{Achievable Information Rates over MIMO Channels}\label{chII-sectionV}
%###################################################################################'
In this section we derive the achievable information rates in the sense of outage rates, associated to a receiver using the decoding rule \eqref{chII-decoder-metric} based on metrics \eqref{chII-final_metric_MIMO} and \eqref{chII-mismached_ML}. 
%---------------------------------------------------------------------------------------------------------------------------
\subsection{Achievable Information Rates  Associated to the Improved Decoder}
%---------------------------------------------------------------------------------------------------------------------------
Assume a given pair of matrices $(\H,\HH)$, characterizing a specific instance of the channel realization and its estimate. We first derive the instantaneous achievable rates $C_{\mc{M}}^\textrm{MIMO}({\H,\HH})$ for MIMO channels $W(\mathbf{y}|\s,\H)=C\mc{N}\big(\H \mb{x},\mb{\Sigma_{0}}\big)$,  associated to a receiver using the derived metric \eqref{chII-final_metric_MIMO}. This is done by using the following Theorem from \cite{merhav-1994}, which provides the general expression for the maximal achievable rate with a given decoding metric.

\begin{theorem}\label{chII-theo_merhav}
For any pair of matrices $({\H,\HH})$, the maximal achievable rate associated to a receiver using a metric $\mc{D}(\s,\mb{y}|\widehat{\H})$ is given by 
\begin{equation}
\label{chII-eq:outage}
C_{\mc{D}}(\H,\HH)=\sup_{P_X\in \mathscr{P}_\Gamma(\mathscr{X})}  \inf_{V_{Y|X} \in \mc{V}(\H,\widehat{\H})} I( P_X,V_{Y|X}),
\end{equation}   
where the mutual information functional
\begin {equation}
\label{chII-eq:I} 
I( P_X,V_{Y|X})=\int\!\!\!\!  \int \!  \log_2 \frac{  V_{Y|X} (\mathbf{y}|\mathbf{x},\Upsilon)} {\int   V_{Y|X} (\mathbf{y}|\mathbf{x}^\prime,\Upsilon) dP_X(\mathbf{x}^\prime)}dP_X(\mathbf{x}) dV_{Y|X} (\mathbf{y}|\mathbf{x},\Upsilon),
\end{equation} 
and $\mathcal{V}(\H,\widehat{\H})$ denotes the set of test channels, i.e., all possibles uncorrelated MIMO channels $V_{Y|X} (\mathbf{y}|\mathbf{x},\Upsilon)=C\mathcal{N}(\Upsilon \mb{x},\mb{\Sigma})$, verifying that\footnote{Our constraint $(c_1)$ is different of that provided in \cite{merhav-1994}, since here the channel noise is i.i.d. and consequently we can only satisfy the equality of the matrix traces and not of the covariance matrices.}
\begin{align*}
&(c_1): tr \big( \mathbb{E}_P  \, \big\{\mathbb{E}_{V} \{\mb{y}\mb{y}^\dag \} \big\} \big) = tr \big(   \mathbb{E}_P \, \big\{ \mathbb{E}_{W}\{ \mb{y}\mb{y}^\dag \}\big\} \big), 
\\  \label{chII-eq:ineqconstr1}
&(c_2): \mathbb{E}_P  \, \Big\{\mathbb{E}_{V} \big\{\mathcal{D}(\s,\mb{y}|\widehat{\H})\big\} \Big\} \leq \mathbb{E}_P \, \Big\{ \mathbb{E}_{W} \big\{\mathcal{D}(\s,\mb{y}|\widehat{\H})\big\}\Big\}.
\end{align*}
%where $\mathcal{H}(\cdot)$ denote the entropy function. 
\end{theorem}

In order to solve the constrained minimization problem in Theorem \eqref{chII-theo_merhav} for our metric $\mc{D}=\mc{D}_\mc{M}$ (expression \eqref{chII-final_metric_MIMO}), we must find the channel $\Upsilon\in \mathbb{C}^{M_R\times M_T}$ and the covariance matrix $\mb{\Sigma}=\mathbb{I}_{M_R} \sigma^2$ defining the test channel $V_{Y|X} (\mathbf{y}|\mathbf{x},\Upsilon)$ that minimizes the relative entropy \eqref{chII-eq:I}. On the other hand, through this paper we assume that the transmitter does not dispose of the channel estimates, and consequently no power control is possible. Thus, we choose the sub-optimal input distribution $P_X=C\mathcal{N}(\mb{0},\mb{\Sigma_P})$ with $\mb{\Sigma_P}= \mathbb{I}_{M_T} \bar{P}$. We first compute the constraint set $\mc{V}(\H,\widehat{\H})$, given by $(c_1)$ and $(c_2)$, and then we factorize matrix $\H$ to solve the minimization problem. Before this, to compute the constraint $(c_2)$, we need the following result (Appendix \ref{chII-Appendix_II}). 

\begin{lemma}\label{chII-elemma}
Let $\mb{A}\in \mathbb{C}^{M_R\times M_T}$ be an arbitrary matrix and $\mb{X}$ be a random vector with pdf $C\mathcal{N}(0,\mb{\Sigma_P})$. For every real positive constants $K_1,K_2>0$, the following equality holds   
\begin{equation}
\mathbb{E}_{\mb{X}}\!\! \left[ \frac{\| \mb{A} \mb{X}  \|^2 + K_1}{\| \mb{X}\|^2+K_2} \right]=\frac{\|\mb{A}\|_F^2}{n+1} + \left ( \frac{K_1}{K_2}-\frac{\|\mb{A}\|_F^2}{n+1} \right) \left(\frac{K_2}{\bar{P}}\right)^{n+1}\!\!\!\!\!  \exp  \left( \frac{K_2}{\bar{P}} \right) \Gamma\left(-n,K_2/\bar{P}\right)\label{chII-result_lemma},
\end{equation}
where $n=M_T-1$ with $n\in \mathbb{N}_+$ and $\Gamma(-n,t)=\displaystyle{\frac{(-1)^n}{n !} \Big[\Gamma(0,t)-\exp(-t)\sum\limits_{i=0}^{n-1}(-1)^i \frac{i !}{t^{i+1}} \Big]}$, $\mb{\Sigma_P}= \mathbb{I}_{M_T} \bar{P}$ and $\Gamma(0,t)= \displaystyle{\int^{+\infty}_{t}\!\!\! u^{-1}\exp(-u)du}$ denotes the exponential integral function.\vspace{1mm}
\end{lemma}

From Lemma \ref{chII-elemma} and some algebra, it is not difficult to show that the constraints require that 
\begin{eqnarray}\vspace{-3mm}
&\mathrm{(c_1)}:&  tr \big(\Upsilon\mb{\Sigma_P} \Upsilon^\dag+ \mb{\Sigma} \big) = tr \big(\H \mb{\Sigma_P} \H^\dag+ \mb{\Sigma_0} \big), \label{chII-c1}\\
&\mathrm{(c_2)}:& \| \Upsilon +a_\mc{M} \widehat{\H} \|^2_F\leq  \| \H +a_\mc{M} \widehat{\H} \|^2_F + \textrm{C},\label{chII-c2} 
\end{eqnarray}
\vspace{-8mm}\,
\begin{eqnarray*} 
a_\mc{M}&=&\delta(\delta\sigma^2_{\mathcal{E}} \bar{P} - \lambda_n \sigma_Z^2)\big[M_T\delta\sigma^2_{\mc{E}}\lambda_n \bar{P}+\lambda_n \sigma_Z^2- \delta\sigma^2_{\mathcal{E}} \bar{P} \big]^{-1},\\
\textrm{C}&=& M_T \lambda_n \big[\|\H\|^2_F-\|  \Upsilon\|^2_F + \bar{P}^{-1}\big(tr(\mb{\Sigma_0})-tr(\mb{\Sigma})\big)\big] \big[1-\frac{\sigma_Z^2}{\delta \bar{P}\sigma^2_{\mathcal{E}}}\lambda_n-M_T \lambda_n  \big]^{-1}, \\
\lambda_n&=&\left(\frac{ \sigma_Z^2}{\delta \bar{P} \sigma^2_{\mathcal{E}} }\right)^n \exp\left(\frac{ \sigma_Z^2}{\delta \bar{P} \sigma^2_{\mathcal{E}} }\right)\Gamma\left(-n,\frac{ \sigma_Z^2}{\delta \bar{P}\sigma^2_{\mathcal{E}} }\right),\,\,\,\,\,\,\textrm{with $n=M_T-1$.}
\end{eqnarray*}
From expression \eqref{chII-c2} and computing the relative entropy, the minimization in \eqref{chII-eq:outage} writes
\begin{equation}
C_{\mc{M}}^\textrm{MIMO}(\H,\widehat{\H})=\left \{ \begin{array}{ll} \min\limits_{\Upsilon} \,\,\,\,\,\, \log_2 \textrm{det}\left(\mathbb{I}_{M_R}+ \Upsilon \mb{\Sigma_P} \Upsilon^\dag \mb{\Sigma}^{-1}\right), \\ \textrm{subject to} \,\,\,\,\,\,  \| \Upsilon +a_\mc{M} \widehat{\H} \|^2_F\leq  \| \H +a_\mc{M} \widehat{\H} \|^2_F + \textrm{C}, \end{array}\right.\label{chII-eq_optb2}
\end{equation}
where $\mb{\Sigma}$ must be chosen such that $tr \big(\Upsilon\mb{\Sigma_P} \Upsilon^\dag+ \mb{\Sigma} \big) = tr \big(\H \mb{\Sigma_P} \H^\dag+ \mb{\Sigma_0} \big)$. In order to obtain a simpler and more tractable expression of \eqref{chII-eq_optb2}, we consider the following decomposition of the matrix $\H=\mathbf{U}\, \textrm{diag}(\underline{\lambda}) \mathbf{V}^\dag$ with $\underline{\lambda}=(\lambda_1,\dots,\lambda_{M_R})^T$. Let $\textrm{diag}(\underline{\mu})$ be a diagonal matrix such that $\textrm{diag}(\underline{\mu})=\mathbf{U}^\dag \Upsilon \mathbf{V}$, whose diagonal values are given by the vector $\underline{\mu}=(\mu_1,\dots,\mu_{M_R})^T$. We define $\widetilde{\H}^\dag=\mathbf{V}^\dag \widehat{\H}^\dag \mathbf{U}$, the vector $\mb{\tilde{h}}^\dag=\textrm{diag}(\widetilde{\H}^\dag)^T$ resulting of its diagonal and let $b_\mc{M}= \| \H + a_\mc{M} \widehat{\H}\|_F^2 - a_\mc{M}^2 (\| \widetilde{\H}\|_F^2-\|  \mb{\tilde{h}}\|^2)$. Using the above definitions and some algebra, the optimization \eqref{chII-eq_optb2} becomes equivalent to 
\begin{equation}
C_{\mc{M}}^\textrm{MIMO}(\H,\widehat{\H})=\left \{ \begin{array}{ll} \min\limits_{\underline{\mu}} \,\,\,\,\,\,  \displaystyle{\sum\limits_{i=1}^{M_R}\log_2 \left(1+\displaystyle{\frac{\bar{P}|\mu_i|^2}{\sigma^2( \underline{\mu})}}\right)}, \\ \textrm{subject to} \,\,\,\,\,\,  \| \underline{\mu} +a_\mc{M}  \mb{\tilde{h}}  \|^2\leq b_\mc{M}, \end{array}\right.\label{chII-final_opt}
\end{equation}
with $\sigma^2( \underline{\mu})=\frac{\bar{P}}{M_R}(\|\underline{\lambda} \|^2-\|\underline{\mu} \|^2)+\sigma_Z^2$. The constraint set in the minimization \eqref{chII-final_opt}, which corresponds to the set of vectors $\{\underline{\mu}\in \mathbb{C}^{M_T\times 1} :\,   \| \underline{\mu} +a_\mc{M}  \mb{\tilde{h}}  \|^2\leq b_\mc{M}\}$, is a closed convex polyhedral set. Thus, the infimun in \eqref{chII-final_opt} is attainable at the extremal of the set given by the equality (cf. \cite{convex-book}). Furthermore, for every vector $\underline{\mu}$ such that $\|\underline{\mu} \|^2 \leq  \|\underline{\lambda}\|^2$, we observe that expression \eqref{chII-final_opt} is a monotonically increasing function of the square norm of $\underline{\mu}$. As a consequence, it is sufficient to find the optimal vector $ \underline{\mu}^{\textrm{opt}}_\mc{M}$ by minimizing the square norm over the constraint set. This becomes a classical minimization problem that can be easily solved by using Lagrange multipliers. The corresponding achievable rates are then presented in the following corollary.

\begin{corollary}\label{chII-corollary_archievable_rates_MIMO}
Given a pair of matrices $(\H,\HH)$ the following information rates can be achieved by a receiver using the decoding rule \eqref{chII-decoder-metric} based on the metric \eqref{chII-final_metric_MIMO}, for uncorrelated MIMO channels,
\begin{equation}
C_{\mc{M}}^\textrm{MIMO}(\H,\widehat{\H})=\log_2 \textrm{det}\left(\mathbb{I}_{M_R}+ \Upsilon_{\textrm{opt}} \mb{\Sigma_P}\Upsilon_{\textrm{opt}}^\dag \sigma^{-2}( \underline{\mu}^{\textrm{opt}}_\mc{M})   \right),\label{chII-acievable_rates_MIMO} 
\end{equation}
where the optimal solution $\Upsilon_{\textrm{opt}}=\mathbf{U}\, \textrm{diag}(\underline{\mu}^{\textrm{opt}}_\mc{M}) \mathbf{V}^\dag$ with 
\begin{equation}
\underline{\mu}^{\textrm{opt}}_\mc{M}= \left\{ \begin{array}{ll}\displaystyle{ \left(\frac{\sqrt{b_\mc{M}}}{\|\mb{\tilde{h}}\|}-|a_\mc{M}|\right)\mb{\widetilde{h}}} & \,\, \textrm{if $b_\mc{M}\geq 0$,} \\
  \underline{0} & \,\, \textrm{otherwise}, \end{array}  \right. \label{chII-solution_mu}
\end{equation}
\end{corollary}
and $\sigma^2(\underline{\mu}^{\textrm{opt}}_\mc{M})=\frac{\bar{P}}{M_R}(\|\underline{\lambda} \|^2-\|\underline{\mu}^{\textrm{opt}}_\mc{M} \|^2)+\sigma_Z^2$.

\subsection{Achievable Information Rates Associated to the Mismatched ML decoder} 

Next, we aim at comparing the achievable rates obtained in \eqref{chII-acievable_rates_MIMO} to those provided by the classical mismatched ML decoder \eqref{chII-mismached_ML}. Following the same steps as above, we can compute the achievable rates associated to the mismatched ML decoder. In this case, the minimization problem writes
\begin{equation}
C_{\textrm{ML}}^\textrm{MIMO}(\H,\widehat{\H})=\left \{ \begin{array}{ll} \min\limits_{\Upsilon} \,\,\,\,\,\, \log_2 \textrm{det}\left(\mathbb{I}_{M_R}+  \Upsilon\mb{\Sigma_P} \Upsilon^\dag \mb{\Sigma}^{-1}\right), \\ \textrm{subject to} \,\,\,\,\,\,  \mathbb{R}e\{tr( \H \mb{\Sigma_P} \widehat{\H}^\dag) \} \leq  \mathbb{R}e \{tr( \Upsilon \mb{\Sigma_P} \widehat{\H}^\dag ) \}, \end{array}\right.\label{chII-eq_optb}
\end{equation}
where $\mb{\Sigma}$ must be chosen such that $tr \big(\Upsilon\mb{\Sigma_P} \Upsilon^\dag+ \mb{\Sigma} \big) = tr \big(\H \mb{\Sigma_P} \H^\dag+ \mb{\Sigma_0} \big)$. The resulting achievable rates are given by
\begin{equation}
C_{\textrm{ML}}^\textrm{MIMO}(\H,\widehat{\H})=\log_2 \textrm{det}\left(\mathbb{I}_{M_R}+ \Upsilon_{\textrm{opt}}\mb{\Sigma_P} \Upsilon_{\textrm{opt}}^\dag \sigma^{-2}(\underline{\mu}^{\textrm{opt}}_\textrm{ML})   \right),\label{chII-acievable_rates_ML} 
\end{equation}
where $\Upsilon_{\textrm{opt}}=\mathbf{U}\, \textrm{diag}(\underline{\mu}^{\textrm{opt}}_\textrm{ML}) \mathbf{V}^\dag$ and 
\begin{eqnarray}
\sigma^2(\underline{\mu}^{\textrm{opt}}_\textrm{ML})&=&\frac{\bar{P}}{M_T}(\|\underline{\lambda} \|^2-\|\underline{\mu}^{\textrm{opt}}_\textrm{ML} \|^2)+\sigma_Z^2,\nonumber\\ 
\underline{\mu}^{\textrm{opt}}_\textrm{ML}&=&\displaystyle{\frac{ \mathbb{R}e\{tr(\Lambda^\dag \tilde{\mb{h}}) \}}{\|\tilde{\mb{h}}\|^2}\tilde{\mb{h}}}.
\end{eqnarray}

\subsection{Estimation-Induced Outage Rates}

Through this section, we have so far considered instantaneous achievable rates over MIMO \eqref{chII-acievable_rates_MIMO} channels. We now provided its associated outage rates, according to the notion of EIO capacity defined in section \ref{chII-brief_review_capa}. In order to compute these outage rates, it is necessary to calculate the outage probability as a function of the outage rate. Given outage rate $R \geq 0$ and channel estimate $\HH$, the outage probability is defined as
\begin{equation}
P_{{\mathcal{M}}}^{\mathrm{out}}(R,\widehat{\mathbf{H}})=\int_{\big\{\mathbf{H}\in \mathbb{C}^{M_R\times M_T} :\, C_{\mathcal{M}}(\mathbf{H},\widehat{\mathbf{H}})<R\big\}} d\psi_{\H|\HH}(\H|\HH), \notag
\end{equation}
then the maximal outage rate for an outage probability $\gamma_{_{QoS}}$ is given by
\begin{equation}
\label{chII-capout}
C_{\mathcal{M}}^{\mathrm{out}}(\gamma_{_{QoS}},\widehat{\mathbf{H}})=\sup_R\big\{R \geq 0: P_{{\mathcal{M}}}^{\mathrm{out}}(R,\widehat{\mathbf{H}})\leq \gamma_{_{QoS}}\big\}. 
\end{equation}
Since this outage rate still depends on the channel estimate, we consider the average over all channel estimates as $\overline{C}_{\mathcal{M}}^{\; \mathrm{out}}(\gamma_{_{QoS}})=\mathbb{E}_{\widehat{\mathbf{H}}}\big\{C_{\mathcal{M}}^{\mathrm{out}}(\gamma_{_{QoS}},\widehat{\mathbf{H}})\big\}$. These achievable rates are upper bounded by the mean outage rates given by the EIO capacity, which provides the maximal outage rate (i.e. maximizing over all possible receiver using the channel estimates), achieved by a theoretical decoder. In our case, this capacity is given by $\overline{C}(\gamma_{_{QoS}})=\mathbb{E}_{\widehat{\mathbf{H}}}\big\{C(\gamma_{_{QoS}},\widehat{\mathbf{H}})\big\}$, where $C(\gamma_{_{QoS}},\widehat{\mathbf{H}})$ can be computed from \eqref{chII-outage-cap} by setting $\theta=\H$ and $\hat{\theta}=\HH$. 

\section{Simulation Results}\label{chII-sectionVI}

In this section we provide numerical results to analyze the performance of a receiver using the decoder \eqref{chII-decoder-metric} based on the metric \eqref{chII-final_metric_MIMO}. We consider uncorrelated Rayleigh fading MIMO channels, assuming that the channel changes for each compound symbol inside a frame  of $L=50$ symbols. This assumption was made because of BICM for interleaver efficiency. The performances are measured in terms of BER and achievable outage rates. The binary information data is encoded by a rate $1/2$ non-recursive non-systematic convolutional (NRNSC) channel code with constraint length $3$ defined in octal form by $(5,7)$. The interleaver is random and operates over the entire frame with size $L \, M_T \log_2(B)$ bits. The symbols belong to a $16$-QAM constellation with either Gray or set-partition labeling. Besides, it is assumed that the average pilot symbol energy is equal to the average data symbol energy.  

\subsection{Bit Error Rate Analysis of BICM Decoding Under Imperfect Channel Estimation}  

\begin{figure}[!htb] 
\centering
\includegraphics[width=5.6in]{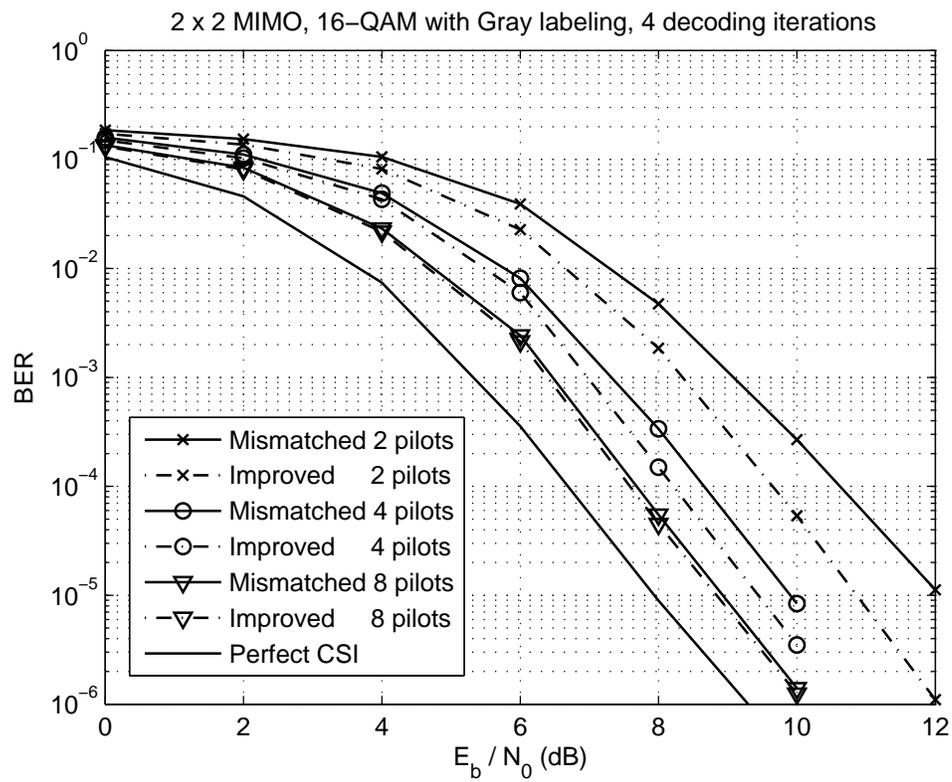} 
\caption{BER performances over $2\times2$ MIMO with Rayleigh fading for various training sequence lengths and Gray labeling.} \label{chII-fig_7}
\end{figure}

\begin{figure}[!htb] 
\centering
\includegraphics[width=5.6in]{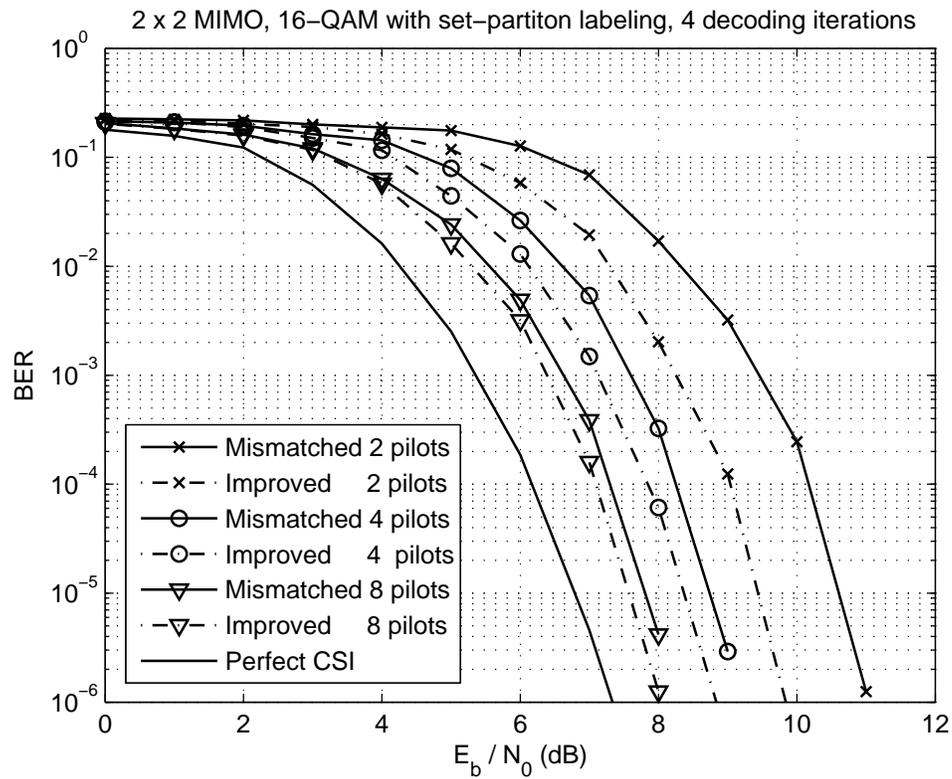} 
\caption{BER performances over $2\times2$ MIMO with Rayleigh fading for various training sequence lengths and set-partition labeling.} \label{chII-fig_7b}
\end{figure}

Here, we compare BER performances between the proposed decoder \eqref{chII-final_metric_MIMO} and the mismatched decoder \eqref{chII-mismached_ML} for BICM decoding (section IV). Fig. \ref{chII-fig_7} and \ref{chII-fig_7b} show, for a $2\times2$ MIMO channel ($M_T=M_R=2$), the increase in the required $E_b/N_0$ caused by decoding with the mismatched ML decoder in presence of CEE. BER obtained with perfect CSIR are also presented for comparison purpose. In this case, we insert $N=2,4$ or $8$ pilots per frame for channel training. At ${\rm BER} = 10^{-4}$ and $N=2$, we observe about $1.4$ dB of SNR gain with set-partition labeling by using the proposed decoder. The performance improvement with set-partition labeling is higher (well served to iterative decoding) than Gray labeling (this is preferred if no iteration is allowed).

We also note that the performance loss of the mismatched receiver with respect to our receiver becomes insignificant for $N\geq8$. This can be explained from \eqref{chII-limit_metric}, since by increasing the number of pilot symbols both decoders coincide. Results show that the decoder under investigation outperforms the mismatched decoder, especially when few are dedicated for training. 

\subsection{Achievable Outage Rates Using the Derived Metric}

Numerical results concerning achievable information rates decoding with the investigated metric over fading MIMO channels are based on Monte Carlo simulations.

\begin{figure}[!htb] 
\centering
\includegraphics[width=5.5in]{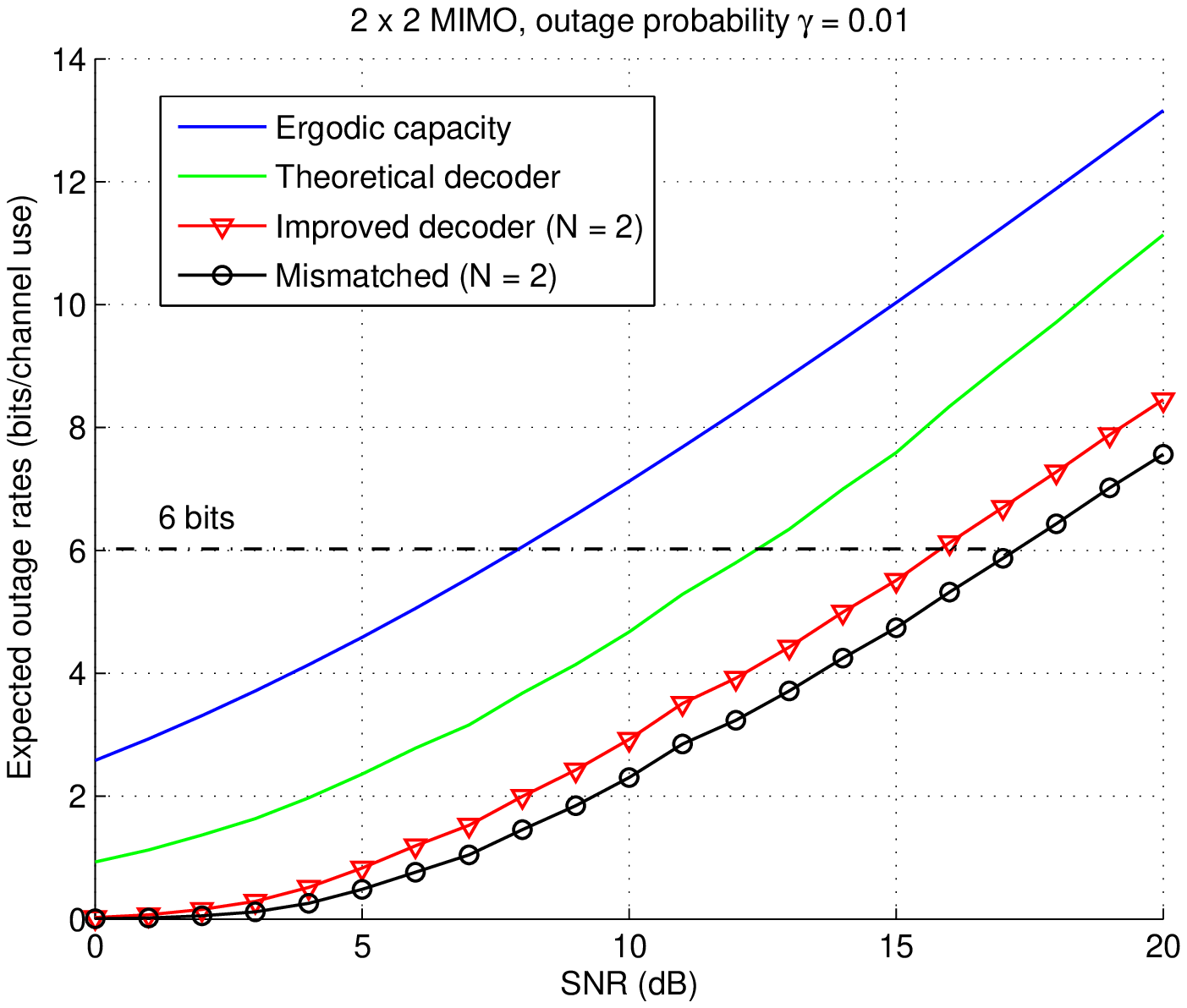} 
\caption{Expected outage rates over $2\times2$ MIMO with Rayleigh fading versus SNR $(N=2)$. }\label{chII-fig_10}
\end{figure}

 Fig. \ref{chII-fig_10} compares average outage rates (in bits per channel use) over all channel estimates, of both mismatched ML decoding (given by expression \eqref{chII-acievable_rates_ML}) and the proposed metric (given by \eqref{chII-acievable_rates_MIMO}) versus the SNR. The $2\times2$ MIMO channel is estimated by sending $N=2$ pilot symbols per frame, and the outage probability has been set to $\gamma_{_{QoS}}=0.01$. For comparison, we also display the upper bound of these rates given by the EIO capacity (obtained by evaluating the expression \eqref{chII-outage-cap}), and the capacity with perfect channel knowledge. It can be observed that the achievable rate using the mismatched ML decoding is about $5$ dB (at a mean outage rate of $6$ bits) of SNR far from the EIO capacity. Whereas, we note that the proposed decoder achieves higher rates for any SNR values and decreases by about $1.5$ dB the aforementioned SNR gap. 
 
\begin{figure}[!htb]   
\centering
\includegraphics[width=5.5in]{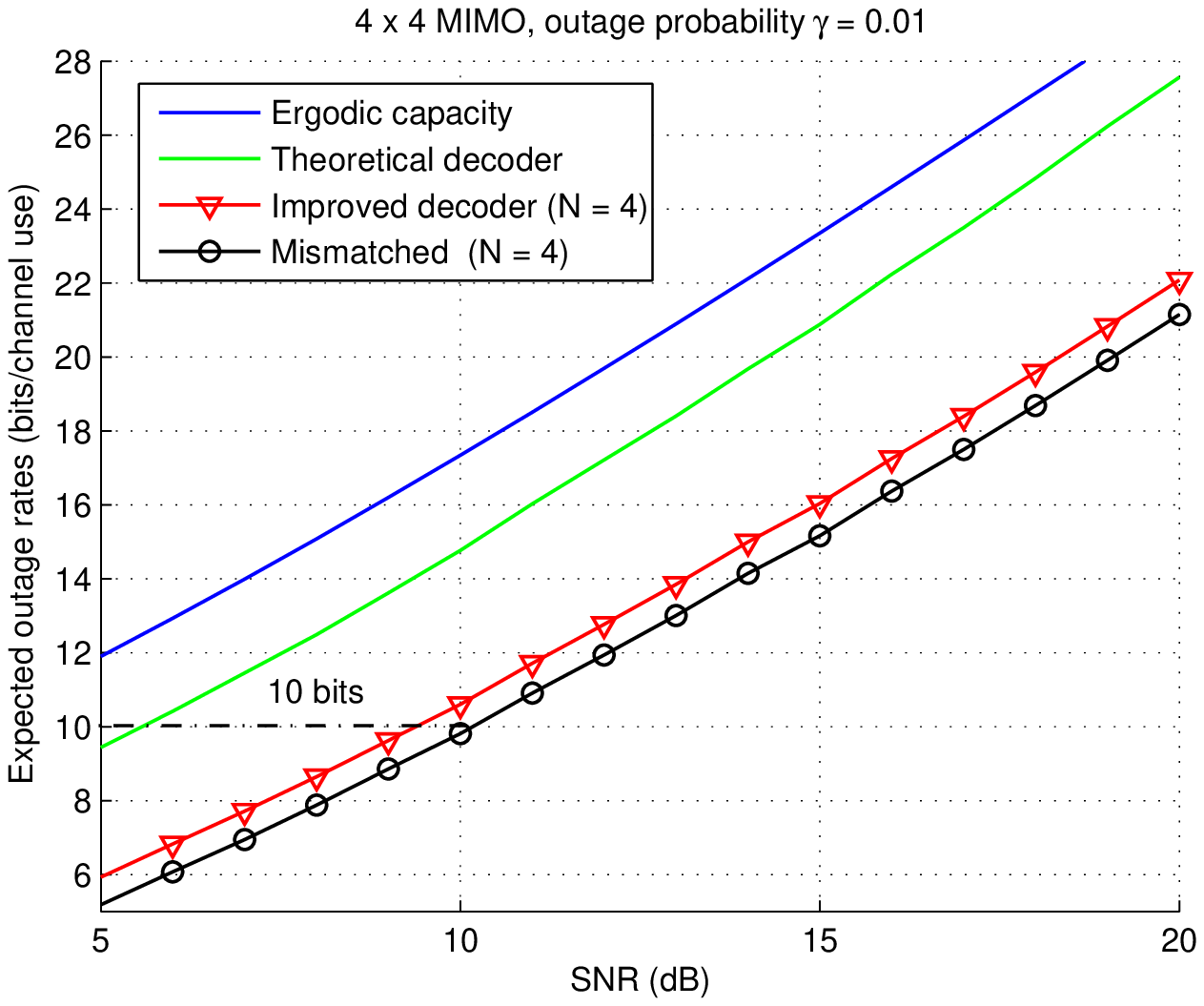} 
\caption{Expected outage rates over $4\times4$ MIMO with Rayleigh fading versus SNR $(N=4)$.} \label{chII-fig_11}
\end{figure}

 Similar plots are shown in Fig. \ref{chII-fig_11} in the case of a $4\times4$ MIMO channel estimated by sending training sequences of length $N=4$. Again, it can be observed that the modified decoder achieves higher rates than the mismatched decoder. However, we note that the performance degradation using the mismatched decoder has decreased to less than $1$ dB (at a mean outage rate of $10$ bits). This observation is a consequence of using orthogonal training sequences that requires $N \geq M_T$ (CEE are reduced by increasing the number of antennas \cite{garg05}). Whereas for $N< M_T$ (using non-orthogonal sequences) the performance degradation will be larger than here.

Note that the achievable rates of the proposed decoder are still about $3$ dB far from the ultimate performance given by the EIO capacity. However, the new metric  provides significative gains in terms of information rates compared to the classical mismatch approach.

\section{Summary}\label{chII-sectionVII}

This paper studied the problem of reception in practical communication systems, when the receiver has only access to noisy estimates of the channel and these estimates are not available at the transmitter. Specifically, we focused on determining the optimal decoder that achieves the EIO capacity of arbitrary memoryless channels under imperfect channel estimation. By using the tools of information theory, we derived a practical decoding metric that minimizes the average of the transmission error probability over all CEE.  This decoder is not optimal in the sense that it cannot achieve the EIO capacity, but it offers improvement performance without introducing any additional decoding complexity.

By using the general decoder, we analyzed the case of uncorrelated fading MIMO channels with ML channel estimation at the decoder and without channel information at the transmitter. Then, we used this metric for iterative BICM decoding of MIMO systems. Moreover, we obtained the maximal achievable rates, using Gaussian codebooks, associated to the proposed decoder and compared these rates to those of the classical mismatched ML decoder. Simulation results indicate that mismatched ML decoding is sub-optimal under short training sequences, in terms of both BER and achievable outage rates, and confirmed the adequacy of the proposed decoder.

Although we showed that the proposed decoder outperforms classical mismatched approaches, the derivation of a practical decoder that maximizes the EIO capacity (over all possible theoretical decoders) under imperfect channel estimation, is still an open problem in its full generality. Nevertheless, other types of decoding metrics incorporating also the outage probability value, have yet to be fully explored.

\appendix\vspace{-2mm}
%\subsection{Metric evaluation}\label{Appendix_I}\vspace{-2mm}

%\section{Auxiliary Proofs}

\subsection{Metric evaluation}\label{chII-Appendix_I}
\begin{theorem}\label{chII-theo_appendix_I}
Let $\mb{H}_i\in \mathbb{C}^{M_R\times M_T}$ ($i=1,2$) be circularly symmetric complex Gaussian random matrices with zero means and full-rank Hermitian covariance matrices $\mb{\Sigma}_{ij}=\mathbb{E}\{(\mb{H})_i(\mb{H})_j^\dag \}$ of the columns $(\mb{H})_i$ of $\mb{H}_i$ (assumed to be the same for all columns) for $i=1,2$. Then the random variable $\mb{H}_1| \mb{H}_2 \sim C\mc{N}(\mu, \mathbb{I}_{M_T}\otimes \mb{\Sigma})$ is a circularly symmetric complex Gaussian with mean $\mu=\mb{\Sigma}_{12}\mb{\Sigma}_{22}^{-1} \mb{H}_2$ and covariance matrix of its columns $\mb{\Sigma}=\mb{\Sigma}_{11}-\mb{\Sigma}_{12} \mb{\Sigma}_{22}^{-1} \mb{\Sigma}_{21}$.
\end{theorem}

From \eqref{chII-pdf_MIMO} and \eqref{chII-apriori-ML-pdf}, by choosing $\mb{\Sigma}_{11}=\mb{\Sigma}_{12}=\mb{\Sigma_{H}}$ and $\mb{\Sigma}_{22}=\mb{\Sigma_{H}}+\mb{\Sigma_{\mc{E}}}$ in Theorem \ref{chII-theo_appendix_I}, we obtain the {\it a posteriori} pdf $\psi_{\mb{H} | \widehat{\mb{H}}_{\textrm{ML}}}(\mb{H} | \widehat{\mb{H}}_{\textrm{ML}})= C \mc{N}\big ( \mb{\Sigma_{\Delta}} \widehat{\mb{H}}_{\textrm{ML}},  \mathbb{I}_{M_T}\otimes \mb{\Sigma_{\Delta}} \mb{\Sigma_{\mc{E}}} \big )$, where $\mb{\Sigma_{\Delta}}=\mb{\Sigma_{H}}(\mb{\Sigma_{\mc{E}}}+\mb{\Sigma_{H}})^{-1}$. In order to evaluate the general expression of the decoding metric \eqref{chII-metric-definition} for fading MIMO channels, we compute the expectation of $\mb{W} (\mb{y} |\mb{x},\mb{H})=C\mc{N}\big( \mb{H} \mb{x}, \mb{\Sigma_{0}} \big)$ over the pdf $\psi_{\mb{H} | \widehat{\mb{H}}_{\textrm{ML}}}(\mb{H} | \widehat{\mb{H}}_{\textrm{ML}})$. To this end, we need the following result (see \cite{book-schwartz}).
\begin{theorem}\label{chII-theo-aux}
For a circularly symmetric complex random vector $\mb{v}\sim C \mc{N} (\mb{\mu},\Pi)$ with mean $\mb{\mu}= \mathbb{E}_\mb{v} \{\mb{v}\}$ and covariance matrix $\mb{\Pi}=\mathbb{E}_\mb{V}\{\mb{v}\mb{v}^\dag\}-\mb{\mu} \mb{\mu}^\dag$, and Hermitian positive definite matrix $\mb{A}$ such that $\mathbb{I}+\mb{\Pi} \mb{A} \succ 0$, we have
\begin{equation}
\mathbb{E}_{\mb{V}}\big[ \exp(-\mb{v}^\dag \mb{A} \mb{v}) \big]=|\mathbb{I}+\mb{\Pi} \mb{A} |^{-1}\exp\big[-\mb{\mu}^\dag \mb{A} (\mathbb{I}+\mb{\Pi} \mb{A})^{-1}\mb{\mu} \big].\label{chII-eq-theo}
\end{equation}
\end{theorem}
From this theorem, we can compute the composite channel $\widetilde{\mb{W}}(\mb{y}|\mb{x},\widehat{\mb{H}})$. Let us define $\mb{v} =\mb{y} - \mb{H}\mb{x}$ such that the conditional pdf of $\mb{v}$ given $(\widehat{\mb{H}}, \mb{x})$ is $\mb{v} | (\widehat{\mb{H}},\mb{x})  \sim  C \mc{N}(\mu,\mb{\Pi})$ with $\mu= \mb{y} - \mb{\Sigma_{\Delta}}\widehat{\mb{H}}\mb{x}$ and $\displaystyle{\mb{\Pi} =\mb{\Sigma_{\Delta}}\mb{\Sigma_{\mc{E}}} \| \mb{x} \|^2 }$. Thus, by defining $\mb{A}=\mb{\Sigma_{0}}^{-1}$ from \eqref{chII-eq-theo} and after some algebra, we obtain $\widetilde{\mb{W}}(\mb{y} |\mb{x},\widehat{\mb{H}})=C\mc{N}\big( \delta \widehat{\mb{H}} \mb{x}, \mb{\Sigma_{0}} + \delta\mb{\Sigma_{\mc{E}}}   \| \mb{x} \|^2 \big)$.

\subsection{Proof of Lemma \ref{chII-elemma}}\label{chII-Appendix_II}
Consider the quadratic expressions $Q_1(\mb{x})= \|\mb{A}\mb{x} \|^2+ K_1$ and $Q_2(\mb{x})=\| \mb{x}\|^2 + K_2$, where $\mb{x}$ is a vector of $M_T$ elements, such that $Q_1,Q_2>0$ \emph{almost surely}. The joint generating function of $Q_1$ and $Q_2$, namely, $M_{Q_1,Q_2}(t_1,t_2)=\Esp_{\mb{X}}\big\{\exp\big( t_1 Q_1(\mb{x}) +t_2 Q_2(\mb{x}) \big)\big \}$. It easy to see that
\begin{equation}
M_{Q_1,Q_2}(t_1,t_2)= \exp \big (t_1 K_1 + t_2 K_2 \big) \big | \mathbb{I}_{M_R}-\big ( t_1  \mb{A}^\dag \mb{A} + t_2 \big ) \mb{\Sigma_P}  \big |^{-1/2}. \label{chII-moment}
\end{equation}
Then from the Gamma integral and setting $t_2=-z$ in \eqref{chII-moment} we have
\begin{equation}
\Esp_{\mb{X}}\big \{Q_1(\mb{x}) Q_2^{-1}(\mb{x}) \big \}=\int\limits_{0}^\infty \Esp_{\mb{x}}\big \{Q_1(\mb{x}) \exp \big[-z Q_2(\mb{X}) \big]  \big\} dz,\label{chII-integral}
\end{equation}
where it is not difficult to show that 
\begin{eqnarray}
\Esp_{\mb{X}}\big \{Q_1(\mb{x}) \exp \big[-z Q_2(\mb{x}) \big]  \big\} &=&\displaystyle{\frac{\partial M_{Q_1,Q_2}(t_1,-z)}{\partial t_1} \Big |_{t_1=0}},\nonumber\\
&=& \big[K_1+2^{-1}tr(\mb{A} \mb{\Sigma_P} \mb{A}^\dag )(1+z \bar{P})^{-1}   \big]\nonumber\\
&&\times(1+z \bar{P})^{-(M_T/2)}\exp \big (-K_2 z \big).
\end{eqnarray}
Finally, by solving the integral in \eqref{chII-integral}, we obtain the expression \eqref{chII-result_lemma}.

\bibliographystyle{ieeetr}
\bibliography{biblio.bib}
%--------------------------------------------------------------------------------------------------------------
  %--------------------------------------------------Figures-------------------------------------------------
%--------------------------------------------------------------------------------------------------------------
\newpage

\end{document}